# Thematic Analysis of 18 Years of PERC Proceedings using Natural Language Processing

**Tor Ole B. Odden[1] and Alessandro Marin[1], Marcos D. Caballero[1,2]**
[1]Center for Computing in Science Education, University of Oslo, 0316 Oslo, Norway
[2]Department of Physics and Astronomy & CREATE for STEM Institute, Michigan State University, East Lansing, 48824 Michigan, USA

## ABSTRACT

We have used an unsupervised machine learning method called Latent Dirichlet Allocation (LDA) to thematically analyze all papers published in the Physics Education Research Conference Proceedings between 2001 and 2018. By looking at co-occurrences of words across the data corpus, this technique has allowed us to identify ten distinct themes or "topics" that have seen varying levels of prevalence in Physics Education Research (PER) over time and to rate the distribution of these topics within each paper. Our analysis suggests that although all identified topics have seen sustained interest over time, PER has also seen several waves of increased interest in certain topics, beginning with initial interest in qualitative, theory-building studies of student understanding, which has given way to a focus on problem solving in the late 2010s. Since 2010 the field has seen a shift towards more sociocultural views of teaching and learning with a particular focus on communities of practice, student identities, and institutional change. Based on these results, we suggest that unsupervised text analysis techniques like LDA may hold promise for providing quantitative, independent, and replicable analyses of educational research literature.

## I. INTRODUCTION: THE CHALLENGES AND OPPORTUNITIES OF A GROWING LITERATURE BASE

There is now an overwhelming amount of scientific research literature, even in fields as relatively young as Physics Education Research (PER). For example, since 2005 Physical Review Physics Education Research (formerly Physical Review Special Topics – Physics Education Research) has published over 700 articles. The Physics Education Research Conference (PERC) Proceedings have published over 1300 papers since 2001. Journals like The Physics Teacher and the American Journal of Physics have published many thousands of articles on education, starting in 1963 and 1933 respectively. There are almost certainly thousands more relevant articles in the science education and general education research literature.

This magnitude of literature presents a significant challenge for researchers in that they will never have enough time to read every new article that comes out. Additionally, other than through targeted literature reviews, it is difficult for researchers to find all of the papers that might be relevant to them. This is doubly the case with new researchers, who have an increasingly long history of research and scholarship with which to familiarize themselves. For this reason, we are more and more reliant on recommendations, whether by colleagues, automated tools such as Google Scholar, or periodic newsletters like those delivered by the American Physical Society.

However, this volume of literature also presents several opportunities. Most scholars have experienced the joy of stumbling on a useful article right when one needs it the most—with so many published articles, there are always likely to be a few undiscovered gems for every reader as long as one knows where to look. At a broader level, this amount of literature indicates that our field has now matured beyond its nascent stages—we have progressed to the point of having sustained several waves of methodological and theoretical development in our decades of existence. Thus, this body of literature can give us insight into the development of our field—where PER has come from and where it is going. However, to perform this kind of analysis we need new methods, as it is an unreasonable proposition to ask a single researcher (or even a research team) to read and summarize all of the articles within the field.

The subfield of machine learning known as natural language processing has, in recent years, developed techniques for this kind of large-scale data mining and text analysis. In this study, we use one such technique, an unsupervised machine learning method know as Latent Dirichlet Allocation (LDA) [1,2], to analyze all of the PERC papers published from the start of the conference proceedings in 2001 through 2018. Using this analysis, we are able to see which themes have been prevalent in the published PER literature and how those themes have changed over time.

The remainder of the paper is organized as follows. In section II, we present an overview of the LDA technique, including its theoretical underpinnings, assumptions, and evaluation methods. We situate this description within a discussion of natural language processing methods intended for a general audience. In section III we describe our specific analysis process start to finish. In section IV we present the results of our LDA model and describe the implications it has for understanding the development of the PER literature over time. Finally, in sections V and VI, we compare these results to other attempts at characterizing the PER literature as a whole [3], unpack the limitations of automated text analysis techniques like LDA, and suggest future areas of study that it might be used to address.

## II. LITERATURE REVIEW: NATURAL LANGUAGE PROCESSING AND LATENT DIRICHLET ALLOCATION

### A. General overview of natural language processing techniques and their utility

We begin with an overview of automated text analysis methods aimed at a general readership. The field of machine learning that focuses on analysis of language and text is known as *natural language processing* (NLP). Broadly speaking, there are two kinds of natural language processing techniques: supervised and unsupervised. *Supervised techniques* take in data that is paired with a set of outcomes and aim to "train" an algorithm to fit, predict, or classify that data. For example, if one has already sorted a set of documents into several different categories, supervised text analysis techniques can be used to "learn" the features of each category in order to predict the categorization of a new, unseen document. The accuracy of this classification can then be tested by running the algorithm on a "holdout set"—a segment of the original dataset that was removed before the algorithm was trained.

In contrast, *unsupervised techniques* do not have a predetermined result—that is to say, users do not specify a desired set of outcomes ahead of time. Instead, they aim to extract latent structure within data that is not known *a priori*: for example, unsupervised NLP can be used to analyzing large amount of texts to determine key themes or features, group texts into thematic clusters, or detect nuances in different styles of speech [4]. Such analyses can, for example, be used to extract insights about trends in social media posts [5], marketing data [6], and scientific research literature [7,8]. These insights and algorithms can then be applied to create classification and recommendation engines such as those used by shopping websites and news publications [9].

Based on the exploratory aims of our research and our data source for this project (articles from the physics education research literature), we have chosen to use an unsupervised natural language processing method. However, because unsupervised NLP methods do not produce particular, pre-determined outcomes, they can be difficult to evaluate. This is especially the case in applications such as textual data mining, in which multiple possible interpretations of a single text can exist simultaneously.

In their introduction to automated text analysis methods, Grimmer and Stewart [4] address these difficulties by presenting four guiding principles for using such methods:

1. ***Text analysis techniques are always based on imprecise models of language***

More concisely, "all quantitative models of language are wrong—but some are useful" (p. 269). In other words, as with all modeling techniques, many assumptions and simplifications must be made in order to even begin the automated text analysis process. Common examples include removing punctuation or certain words, disregarding word order, and simplifying the many nuances of language. In this way, automated text analysis methods are epistemologically similar to the simplified "toy models" that are regularly used in both physics and physics education research [10].

2. ***Text analysis is used to get insights into data, but cannot replace careful analysis by humans.***

Unlike supervised learning methods, in which one can directly compute the accuracy of one's results, unsupervised methods have no "ground truth" to check against. And, as previously mentioned, it is common for a single text to have multiple possible interpretations for different readers. At the end of the day, humans are always the most reliable evaluators of the results and meaning of automated text analyses. Thus, these methods can amplify what humans are able to do, but cannot replace human evaluation and interpretation.

3. ***There is no single best method for text analysis***

There are many different approaches to automated text analysis, both supervised and unsupervised. Each method has certain strengths and weaknesses—that is, each reveals certain aspects of text data, while hiding others. Particular methods, or models, must be chosen based on the research questions under investigation.

4. ***Validation is very important***

Because unsupervised machine learning methods cannot be directly evaluated on their accuracy, it is important that researchers spend the time evaluating and validating their results using a variety of different methods. Grimmer and Stuart suggest combining "experimental, substantive, and statistical evidence to demonstrate that the measures are as conceptually valid as measures from an equivalent supervised model" (p. 271). We describe several approaches for validation of our chosen technique, Latent Dirichlet Allocation, in section II.D.

With these guidelines in mind, we are aiming to use automated text analysis to understand how PER has conceptualized and studied teaching and learning over time, based on the publications produced over its history. To perform this analysis, we have chosen to use an unsupervised text analysis technique called Latent Dirichlet Allocation.

### B. Latent Dirichlet Allocation

Latent Dirichlet Allocation is a generative probabilistic modeling technique that tries to extract the latent or hidden themes from a set of texts or documents [1]. It assumes that documents are mixtures of a certain number of latent topics, where each topic is a mixture of words. The goal of LDA is to take a group of texts and extract these latent "topics" from them, in the form of groups or sets frequently co-occurring words. The model requires a user to specify the number of topics (K) and a numeric parameter for how "mixed" these topics should be ($\alpha$) in advance. Thereafter the algorithm iteratively "learns" the distribution of words in each topic, and the distribution of those topics in each document. Finally, the user can take these groups of words and, assuming they are recognizably distinct from one another, apply meaning to them and give them a topic name or label.

As an illustrative example, Figure 1 shows an LDA model that we trained on roughly 9000 Huffington Post headlines [11] from 2012 to 2018. We chose to train the model on three distinct subsets of headlines—those that had been previously classified (by a human coder) under the headings of "sports", "religion", or "technology"—in order to see whether the model could produce topics that fit with these three pre-established themes. Based on the results, it seems to have been successful; when given K = 3 topics, the model was able to produce recognizable sets of words that fit with each of the expected themes—for example, *Sports* includes the words "game," "nfl," "win," "player," and "fan," while *Religion* includes words such as "pope_francis," "christian," "church," and "god." Notice, however, that the model is not completely coherent from a human-interpretability standpoint, as it has associated the word "muslim" (which clearly belongs with the religion topic) most strongly with the *Tech* topic.

The model was also able to classify headlines according to their topic mixtures, and examples are shown that were classified (to within a tolerance of 5%) as purely one topic, 50/50 mixtures of two topics, and equal mixtures of all three topics: for example, one headline that was classified as being approximately 50% Religion and 50% Technology was the 2016 headline "Mark Zuckerberg Met with Pope Francis and Gave Him a Drone" [12], shown at the bottom of the figure.

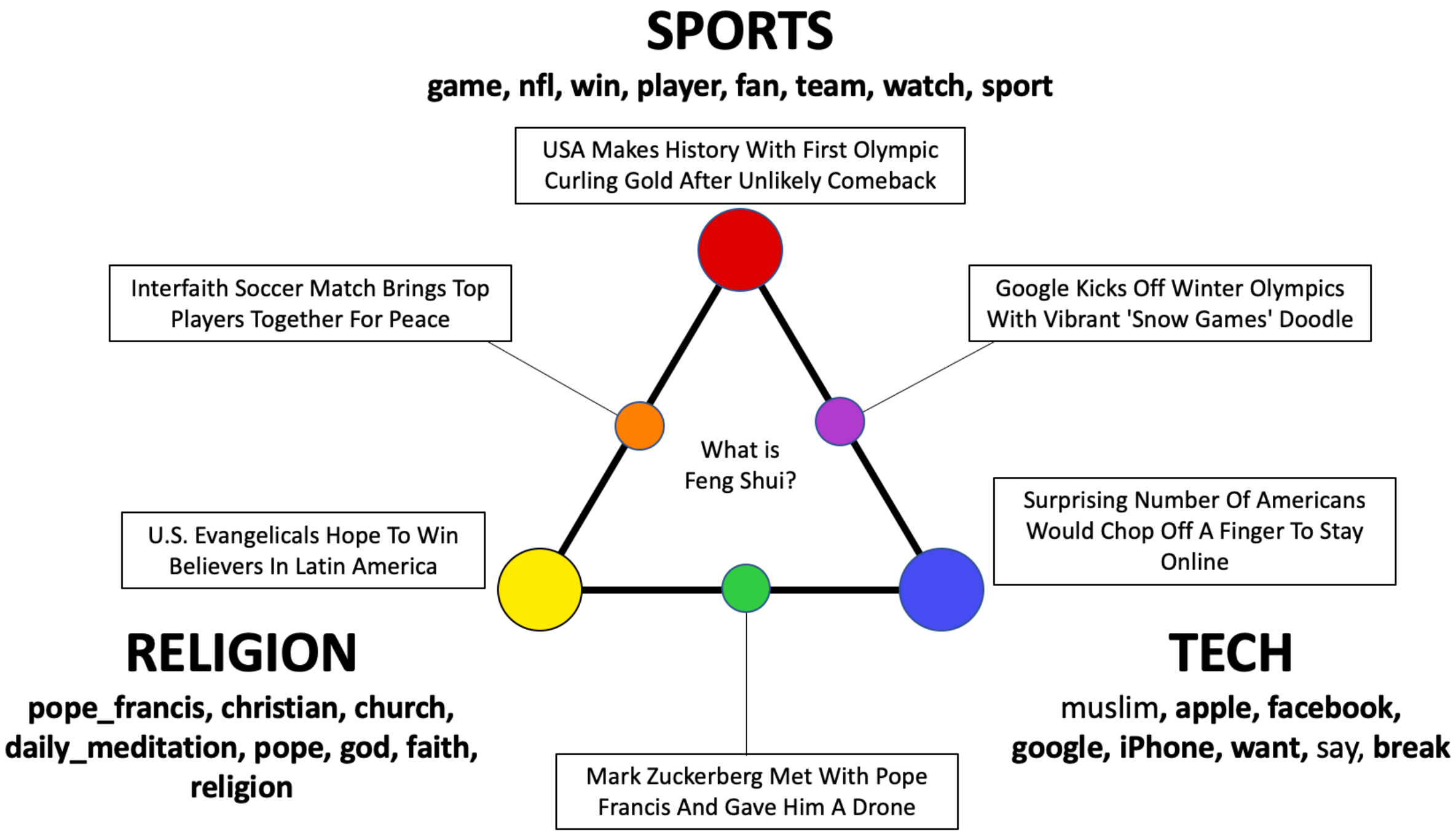

**Figure 1:** Example results from an LDA model trained on a dataset of roughly 9000 Huffington Post headlines from 2012-2018 [11].

LDA derives its topics by assuming a simplified, probabilistic model[1] for how texts are written, similar in nature to how physicists use simplified "toy models" of systems to make physical predications (i.e., neglecting friction or inter-atomic forces). Within this model of the writing process, each "topic" is conceptualized as a weighted probabilistic distribution of all of the possible words that could be written, with the weight on each word determining how central that word is to the topic (that is, how likely it is to be used in texts about that topic). Thus, every topic is assumed to include every possible word in the corpus, but weighted differently depending on the nature of the topic. For example, in the model shown in Figure 1 the Tech topic would rate the probability of choosing technology-focused words such as "apple," "facebook," and "google" very highly, and non-tech related words like "pope_francis" much lower.

According to this model, all topics are assumed to be specified in advance and known by every author. When an author sits down to write a text, they do so also knowing the exact number of words they will write and the relative mixture of topics they plan to use in their document (which is known as the *document-topic distribution*). Then, they generate each word as follows: first, they randomly pick the topic that word will be based on by drawing from their document-topic distribution—one can imagine this as rolling an unfair die, where each side corresponds to one of the possible topics in the document. In the LDA model shown in Figure 1 this would involve randomly choosing one of the three main topics, "Sports," "Religion," or "Tech". Once this is done, they randomly draw a word from that particular topic's distribution

[1] Although we have described this model in qualitative terms here, we provide a more mathematically detailed description of this model in Appendix A. Interested readers are also encouraged to consult the descriptions in Refs [1,8,19]

over words (e.g., "pope_francis" or "apple"). They then repeat this process until all of the words have been chosen in their document.

Using this model, LDA is then able to reverse-engineer the topics in a set of texts, estimating these topic distributions based on the co-occurrence of words in each document. That is, starting with a random initialization of document-topic and word-topic distributions, the LDA algorithm[2] iteratively examines the words that occur in each document and updates these distributions until they reach stability, at which point (if all goes well) each topic is primarily composed of words that frequently occur together. Note, however, that both the model of writing assumed by LDA and this iterative algorithm disregard word order, which is known in the field of NLP as the "bag of words" approach [4].

Although it is based on a very simplified model of writing, LDA has been successfully used in numerous applications. Since its proposal by Blei *et al.* in 2003 [1], LDA has been used in the business world for spam filtering [13], analyzing large software systems [14], and analyzing brand data for marketing [6]. The New York Times has used LDA to create a recommendation engine for its articles which constantly updates as new articles are written and which takes into account reader histories and preferences [9]. In the academic sphere, LDA has been used to analyze political texts [4], dissertations in library science [15] and scientific research articles [8,16]. Over time the technique has been refined, leading to the development of an online algorithm based on variational inference that make it possible to analyze streams of data [2], incorporation of other latent variables like sentiment analysis [6], and development of a supervised variant of the algorithm [17]. However, to our knowledge it has not yet been widely deployed within educational research domains, much less discipline-based educational research communities like PER.

We feel that this technique holds promise because it has the potential to reveal interesting patterns in how our field has conceptualized and re-conceptualized approaches or theories central to the study of teaching and learning. Thus, our research questions are as follows:

1. Based on an LDA analysis of a selection of the Physics Education research literature, what approaches and theories have been prevalent over the last two decades?

2. How has the prevalence of these approaches and theories evolved over time?

## C. LDA Inputs and Outputs

In practice, the inputs and outputs of the LDA algorithm are matrices in which each row is a probability distribution, as shown in Figure 2. The input is a matrix where the rows correspond to documents, and the columns are all the words across the entire corpus (size $D$ x $V$). Each entry corresponds to the word count, or the number of times a word occurs in the corresponding document. The outputs are two factorized matrices, $\theta_D$ and $\beta_K$, that split the input using a set of topics $T_{1:K}$. $\theta_D$ is the *document-topic matrix*, which has rows corresponding to documents and columns corresponding to topics (dimensions $D$ x $K$). The entries are amount of each topic the model predicts is present in each document. $\beta_K$ is the *topic-word matrix*, where each row is a topic and each column is a single word from the corpus (dimensions $K$ x $V$). The entries are the likelihood of a particular word being drawn, given a particular topic. Thus, LDA can also be thought of as a kind of probabilistic matrix factorization [2].

[2] Those interested in how the LDA algorithm derives these distributions are encouraged to consult the descriptions of the algorithms in Blei et al. [1] and Hoffman et al. [2]

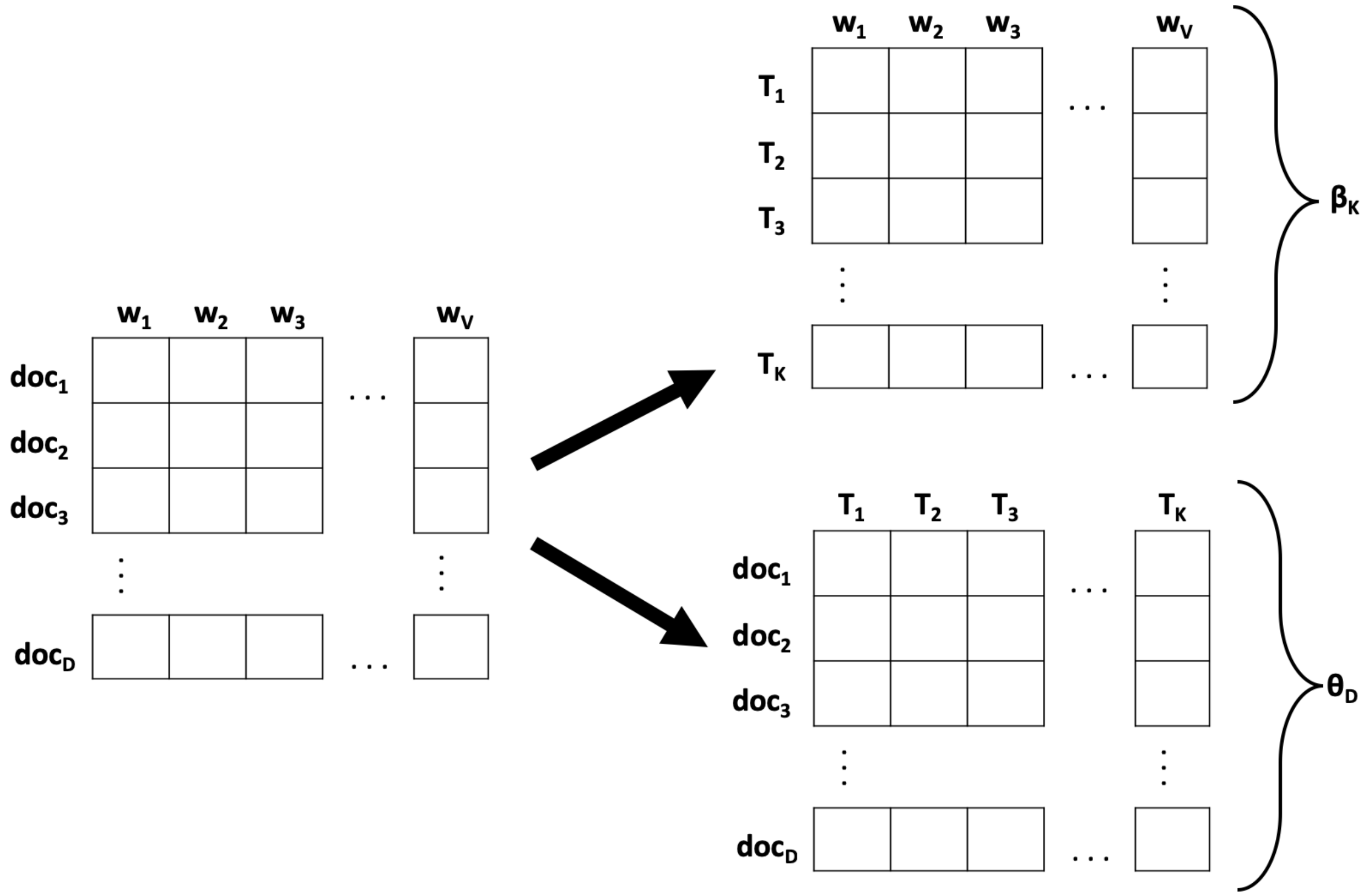

**Figure 2:** Probabilistic matrix factorization interpretation of LDA

Note, however, that in both of the output matrices, each row is a probability distribution across different discrete outcomes (choice of topics or choice of words) so each row will add up to 1. The values in a particular entry will also vary considerably. For example, in the topic-word matrix $\beta_K$, if one had a corpus containing 4000 words and every word in a topic $T_k$ was equally likely to be chosen, the values in that row would uniformly be 1/4000 = 2.5e-4. However, in actual models, most topics have a large number of words that are rated very low, on the order of 1e-6 or 1e-7, while a select set of words will be rated highly, on the order of 1e-2 or even higher. It is therefore common, when presenting topics, to pick the 5-10 most likely words for a particular topic and use those as the descriptor for that topic. This is how we chose the words for the topics shown in Figure 1.

As can be seen from this matrix factorization view, LDA is based on several assumptions about the input dataset. First, as previously mentioned, it assumes that word order does not matter, only word counts in each document. Although this is a major assumption, which disregards much of the nuance of language, it is commonly made in natural language processing studies. It is also in line with Grimmer and Stewart's [4] first principle of text analysis, "all quantitative models of language are wrong, but some are useful." Second, LDA assumes that all texts will necessarily be composed of a mixture of several topics—in other words, this is a "mixed membership" model, rather than a model that requires that each document only be classified into a single category. So, for example, given the 3 topics from Figure 1, "sports", "religion", and "tech", one document might be classified as 20% sports, 50% religion, and 30% tech; another document might be 90% religion, 6% sports, and 4% tech; and so on. However, we feel that this is a major strength of the model when trying to analyze a community like physics education, which studies a fairly narrow set of phenomena and frequently mixes, borrows, and exchanges ideas, theories, and methods (as opposed to, for example, the diverse set of disciplines and methods present in a publication venue like *Science*). Third, LDA assumes that words that are central to a specific topic are more likely to be used more frequently in documents about that topic, and that the most highly-weighted words within that topic will tend to frequently co-occur within documents. This is known as the distributional hypothesis of linguistics [18]. So, for example, if the general topic of a text is dog ownership, the words "dog," "bark," "leash," and "kennel" are likely to be used more often and co-occur more frequently than the words "truck," "laptop," "spaghetti," or "particle accelerator."

## D. LDA Model Evaluation

In the end, the goal of this type of analysis is to produce a single, trained model that can be used to analyze and make claims about both the textual data it was trained on and future data not yet seen. However, as previously mentioned, because LDA is an unsupervised method it requires multiple levels and types of validation in order to ensure that the model results make sense and that one is not arbitrarily choosing a model whose results deviate significantly from the norm. In other words, due to the unsupervised nature of LDA it can be difficult to gauge how one's choice of input parameters will affect the eventual model results. So, in practice, researchers frequently use a variety of evaluation methods, individually or in combination, to compare the quality of LDA models generated under different conditions [4,19,20]. There are several such approaches discussed in the literature:

### 1. *Coherence of topics*

Coherence is essentially a measure of how well a topic "hangs together" based on the tendency of the top words in the topic to co-occur. More technically, it can be thought of as a measure of the degree of semantic similarity between high-scoring words in a topic—that is, are those words being used together and/or in a similar way? This is based on the "distributional hypothesis" of linguistics, that says that words with similar meanings tend to occur in similar contexts [18,19]. Although there are many ways of evaluating coherence, the most commonly-used approach creates a "sliding window" across documents, capturing sections of text and looking at the co-occurrences of words from a particular topic within that window [21]. Words that tend to occur together produce a higher-scoring (that is, more coherent) topic. This score, called the "Coherence Value" or $C_V$, is then normalized to a scale of 0-1, with a higher value indicating a more coherent topic.

### 2. *Perplexity of topic models*

Perplexity is a statistical measure of the log probability of new data (unseen test documents) fitting into the model, based on the geometric mean of the per-word likelihood. Essentially, perplexity is a measure of how well the model fits, represents, or reproduces the statistics of a new set of unseen data, such as Wikipedia articles. However, perplexity has been found to correlate poorly with human interpretability [22].

### *3. Face validity*

Face validity involves using human domain experts to evaluate the quality of produced topics and to make judgments on their consistency and representativeness [23]. In other words, checks of face validity require someone who is familiar with the body of text under analysis to make a judgment about how well the topics fit with their knowledge of the target domain. This is especially helpful when the researchers doing the analysis are not themselves experts in the analyzed literature base [19].

### *4. Comparisons with qualitative methods*

One can also use standard qualitative research methods to evaluate the distinctiveness and interpretability of topics. This can, for example, involve comparing model outputs to the results of human coding using researcher-generated coding schemes [4]. In contrast to face validity checks, these kinds of methods do not necessarily have to involve domain experts—for example, they can involve "intrusion tasks" [22] in which extraneous "intruder" words are inserted into topics and human raters must try to determine which of the words is the intruder.

For our study, we used two of these validation methods. Our primary metric for evaluating our models was to compute the average coherence score based the top 20 words in each topic. We supplemented this metric with occasional checks on the face validity of our results, as two of the authors of this study are members of the PER community who are familiar with the general literature and so were able to provide some judgments as to the face validity of the topics. We also showed preliminary results at various points in the analysis to several other members of the community, from a variety of backgrounds, to make sure that our emerging results made sense. We did not use either perplexity or qualitative methods in this study, but we see this as a rich area of research for future projects.

## III. METHODS

Our primary tool in this analysis was Gensim, an open-source and widely-used Python package for doing natural language processing and topic modeling using machine learning. Gensim includes an LDA algorithm based on variational inference [2], along with other useful functionality for processing and filtering text data, and certain additional packages for visualization. Along with Gensim we also used the data analysis package pandas, the Natural Language Tool Kit (NLTK), and standard scientific computing libraries such as numpy, scipy, and matplotlib.

Our research approach consisted of 4 stages, as shown in Figure 3. We elaborate on each of these stages below.

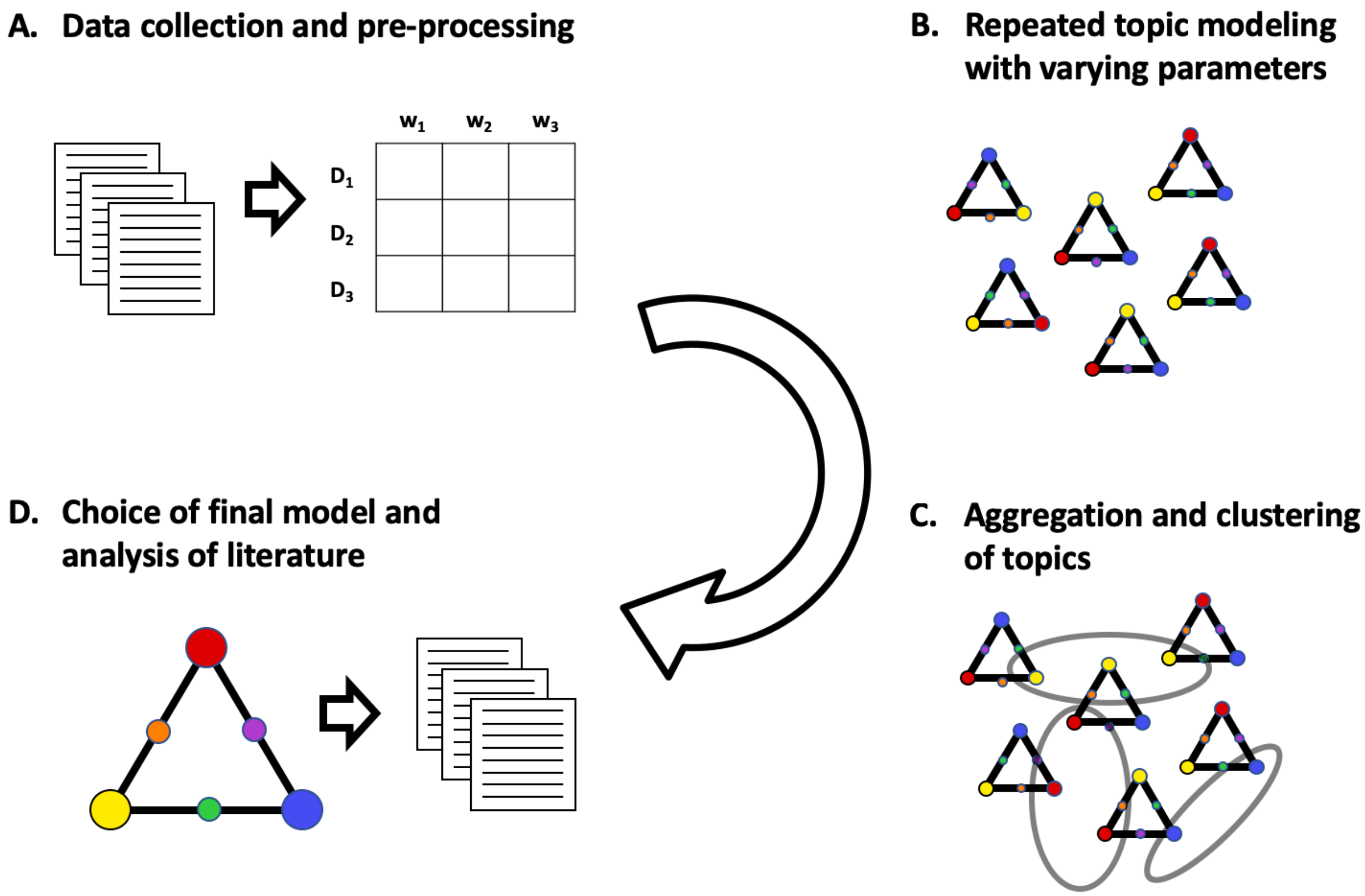

**Figure 3:** Flowchart diagram of methods: A. Data collection and pre-processing; B. Repeated topic modeling across filtering parameters, number of topics, and α values; C. Aggregation and clustering of topics; D. Analysis of literature with final model.

### A. Data collection and pre-processing

Our initial data collection involved downloading all available PDFs of the PERC proceedings from 2001-2018, which are available online through an open-access license. In total, this resulted in 1302 individual papers. We chose to use PERC proceedings because they provided a useful snapshot of the research being done in the PER community at an intermediate level of development. In other words, PERC proceedings are peer-reviewed, which introduces a minimum threshold of research progress in order to be accepted. However, PERC papers are also understood to reflect work-in-progress, which may or may not lead to publication in a full-length format later on. At 4 pages each, PERC papers are significantly easier to write than a full-length research article, and there is no charge to publish in the proceedings. This, we would argue, makes them a more representative sample of the work that has been done than analysis of full-length papers, for example from PRPER.

After downloading all of the papers, we used the package PyPDF2 to scrape the text into a file. We then performed a series of data-cleaning steps, as is standard for text-based machine learning research [24]. First, we applied minor corrections and filtering for cases where papers were included twice (for example, due to corrections) or where the text of a paper had been corrupted. We then removed the references and acknowledgments sections from the texts. We then filtered the text itself by removing numbers/symbols and punctuation, lower-casing all words, reconnecting words that had been broken across lines, removing section headers (i.e., "Introduction", "Conclusion"), and removing "stopwords" (common words that do not carry any specialized meaning, such as "the", "a", "and", "which", or "of"). Finally, we split the resulting blocks of text into lists of individual words, or "tokens."

After this initial round of data cleaning we used a method from NLTK to lemmatize the words in our dataset. *Lemmatization* is the process of matching words with the same or similar meaning in order to reduce the size and complexity of a dataset. For example, the words "test," "tested," "testing," and "tests" all refer to the same essential activity, but without additional processing LDA would treat them as separate words. A lemmatizer reduces such words to their bases or "stems"— for example, it would reduce all instances of the above-mentioned test-related words to the same base word, "test." This kind of data reduction can have the unintended side-effect of reducing the amount of granularity and detail in the data—for example, "physics" and "physical" might be reduced to the same base word, "physic", even though they tend to be used in different ways. To offset this limitation, we included a part-of-speech tagger (taken from the NLTK library) in our lemmatization function, which helps to distinguish between cases in which a word was used as a noun, verb, adjective, or adverb and match words across different parts of speech. Thus, our final set of words includes, for example, both "gravitational" and "gravity" (distinguishing similar words used in different parts of speech) but has reduced all instances of "answer," "answers," and "answering" to the base-word "answer."

Next, we used an algorithm to find and create bi-grams, which are pairs of words that are frequently associated and may carry additional meaning together—for example, "problem solving," "sense making," and "high school." These were then joined together into a single word with an underscore, such as "high_school."

Finally, we took the resulting lists of words, and reduced them into a "bag of words" matrix, as described above. This left us with a matrix in which each row represented a document, each column a single word, and each entry the number of times that word appeared in that document. After pre-processing, we were left with 29026 words and bi-grams.

### B. LDA modeling and grid search across hyperparameters

The next step involved filtering out the words that occurred very frequently and very rarely in the data. This is a critical step to topic modeling techniques like LDA, because words that occur too frequently provide little insight into the data and can wash out interesting trends, while those that occur seldom (such as names of people or institutions) can significantly increase processing time. To determine the optimal values of these filtering parameters, we created a variety of LDA models with different combinations of the filtering parameters and evaluated them based on the spread and averages of the models' coherence scores, aiming to find the combination that maximized coherence. Based on these tests, we chose the following filtering parameters:

1. ***Removal of all words that occurred in more than 55% of the documents (frequent words).***

The 5 most frequent words in our dataset included "student", "physic", "question", "course", and "learn." After creating a variety of models with different filtering thresholds for the most frequent words and evaluating their average coherence scores, we set this threshold to 55%. This removed the previously-mentioned top 5 words, plus 97 others, for a total of 103 words. These words, along with a graph of the frequency of the top 20 words in the corpus, are shown in Figures 4a and 4b.

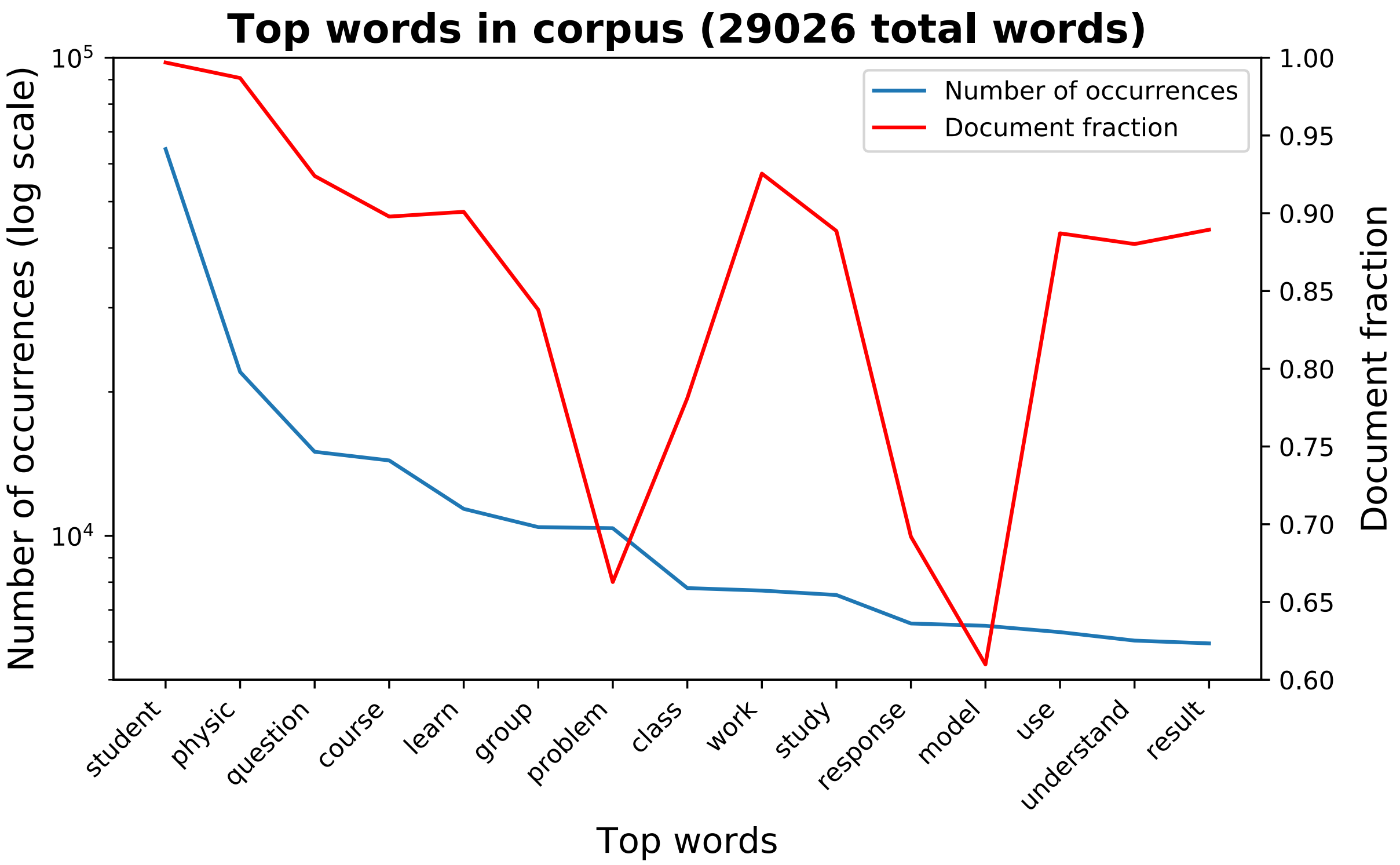

**Figure 4a:** Frequency of 20 most common words in the data corpus. For each word, both the overall number of occurrences and the fraction of documents containing that word are shown.

| | | | | |
|---|---|---|---|---|
| *able* | *design* | *include* | *order* | *similar* |
| *analysis* | *determine* | *indicate* | *paper* | *small* |
| *analyze* | *develop* | *individual* | *particular* | *student* |
| *answer* | *developed* | *instruction* | *physic* | *study* |
| *approach* | *difference* | *instructor* | *point* | *suggest* |
| *ask* | *different* | *introductory* | *possible* | *support* |
| *base* | *discussion* | *involve* | *present* | *table* |
| *begin* | *education* | *keywords* | *problem* | *take* |
| *case* | *example* | *know* | *process* | *teach* |
| *change* | *experience* | *knowledge* | *provide* | *term* |
| *class* | *explain* | *large* | *question* | *test* |
| *common* | *focus* | *learn* | *reason* | *think* |
| *compare* | *follow* | *level* | *related* | *time* |
| *complete* | *give* | *like* | *require* | *type* |
| *concept* | *goal* | *make* | *research* | *understand* |
| *conceptual* | *group* | *mean* | *response* | *university* |
| *consider* | *help* | *model* | *result* | *use* |
| *context* | *high* | *need* | *science* | *way* |
| *course* | *idea* | *new* | *second* | *work* |
| *data* | *identify* | *note* | *set* | |
| *department* | *important* | *number* | *show* | |

**Figure 4b:** Alphabetical list of most frequent words in data corpus, removed using frequency filter of 55%

We have chosen to remove these words because our goal in this study is to reveal interesting differences and structures in the PERC proceedings data. However, in order to see such differences we needed to remove words that are widely used within the community but may carry different meanings for different research groups based on their methods and theoretical commitments. Most research groups study topics that involve “physics,” “learning,” and “courses” to some degree (see Figure 4a), so it would be no great finding to see that these topics are widespread within PER. Additionally, words such as “model” can be used in multiple different contexts—for example by researchers studying how students build mathematical models [25], computational models [26], generalized mental models [27], use model-based reasoning [28], or who are defining “toy models” of student cognition [10]. Words with such wide-ranging meanings can cause problems for topic analysis methods like LDA. However, because these words are so prevalent, they are likely to be caught by frequency-based filters like the one used above.

2. ***Removal of all words that occurred less than 15 times across all papers in the dataset (rare words).***

The filtering for rare words removed many more, on the order of 24,500 words. Although this is a significant amount of the dataset, we made an assumption (common to work in NLP, e.g., [24]) that these words were unlikely to be central to any specific research topic, because they occurred so rarely. This cut allowed us to both increase processing speed, due to the reduced dimensionality of the resulting dataset, and also eliminate words that would likely only contribute noise and blur interesting results.

After applying these two filters, we had reduced our dataset from 29,026 unique words and bigrams to 4645.

In addition to testing the effects of different combinations of these two filtering thresholds, we also computed a variety of models to test the effects of varying our two primary hyperparameters, the numbers of topics (K) and relative mixtures of topics[3] (α), evaluating the results based on the spread and average value of the models' coherence scores. After these tests, we were unable to distinguish any major differences based on α values, so in the next stage of the project (clustering and aggregation) we created datasets with a mixture of α values. We were, however, able to see differences in model performance by varying the number of topics, K, that the model produced. In order to choose the number of topics to include in our final model, we used an "elbow method" [19], which involves training a large number of models with varying numbers of topics, plotting the average coherence score vs. number of topics, and then choosing topic number that appears at or above a leveling-off point in the graph. Since we used a mixture of α values, we generated elbow plots for each of α = 1, 5, 7.5 10, and 12.5, and examined each of these plots separately. Three of these plots shown in Figure 5, with α values of 1, 10, and 12.5; one can see that the average coherence scores in each graph rises with increasing numbers of topics until the model includes 8-10 topics, at which point they levels off, indicating a point of diminishing returns. Based on these "elbows" for the different plots and several "face validity" checks of sample models, we chose a value of K = 10 topics for our final model.

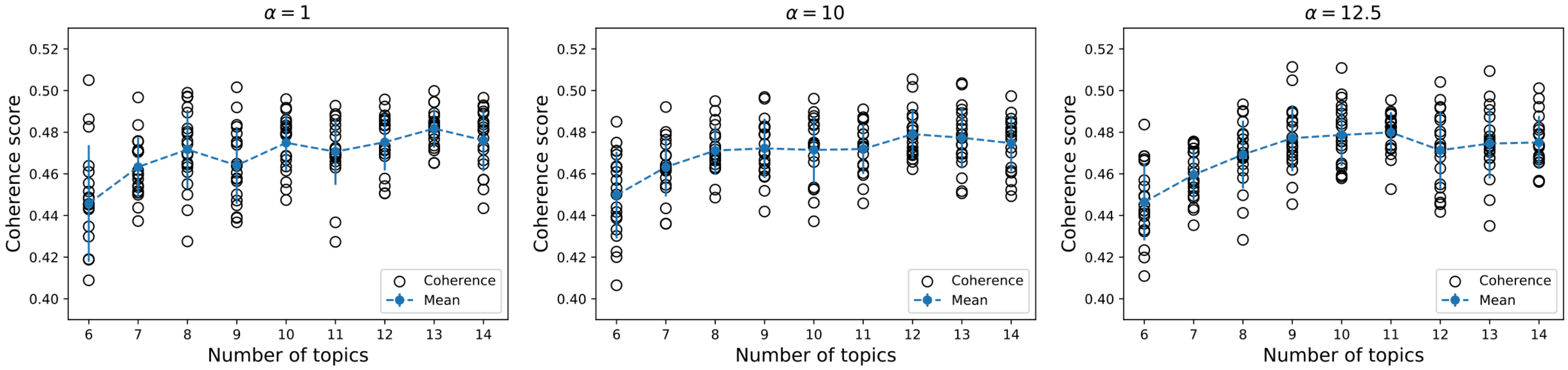

**Figure 5:** Graphs of elbow plot with different document-topic mixtures (α values)

It is worth noting that coherence values, which we used as our primary metric for evaluating topics, vary between 0-1, with higher values indicating more coherent topics. In most of our models we observed coherence scores between 0.4 and 0.52. Although these scores are slightly lower than values achieved by other studies of scientific literature through LDA (e.g., [19]) we suspect this is a function of the relatively small amount of text included in our analysis. Syed and Spruit [19] have shown that average coherence scores seem to increase with additional documents (up to a point) in their analysis of 15,000 articles from the field of fisheries and aquatic science. Thus, considering our corpus consisted of ~1300 4-page papers, we find these coherence values to be acceptable.

[3] For more information on the meaning of the α value, see Appendix A.

### C. Aggregation of topics, cluster analysis, and choice of final model

After determining our model parameters, we encountered a significant challenge with the modeling process: because the LDA algorithm randomly initializes both the topic-word matrix $\beta_K$ and the document-topic matrix $\theta_D$, topics tend to change from run to run depending on the random initial state of the model. In other words, a single model is replicable, as long as it uses the same initialization each time (which is usually provided in the form of an integer-valued random seed). However, if one were train two models, both using the same dataset and parameters but with different random initializations, the models could return significantly different sets of topics and document categorizations. One can see evidence of this issue from the graphs shown in Figure 5: each datapoint in that figure is a model with a different random initialization, resulting a slightly different set of topics and varying coherence values. One can think of this variation as being analogous to the LDA function finding local minima in the complex, multi-dimensional word space defined by our corpus. However, such random effects are a major obstacle if one is trying to choose a particular model to use, because different models could potentially produce different sets of claims.

These issues are to our knowledge not well addressed in the LDA literature. Some studies acknowledge them, while many others ignore them. For example, Roberts et al. [23] state that "Blei (2012) provides an excellent overview of LDA and related models but does not mention the issue of local optima at all. The original paper introducing LDA, mentions local optima only in passing to warn against degenerate initializations" (p. 9). Those studies that do confront the issue have used techniques such as aggregating coherence scores across multiple runs and then choosing the single model with the highest coherence score [19], comparing the topics of multiple generated models to those of a single "reference model" [23], clustering and aggregating individual topics across multiple runs [5], and using other methods such as non-negative matrix factorization that can be easily aggregated [29]. However, although these techniques are sufficient for generating stable sets of topics we found that they did not allow us to perform our desired analyses, such as computing topic prevalence over time.

For this reason, in our study we stabilized our model using a clustering and aggregation technique inspired by Mantyla et al. [30] and Mehta et al. [31]. After creating and examining multiple models with different random initializations, we noted that although topics tended to change from run to run, they seemed to circulate around certain recurring themes. For example, one model might have a topic focused around the term "labs" and another topic based on the word "tutorial", while another model would have a blended topic incorporating both "labs" and "tutorial." Based on this observation, and in line with other documented approaches in the literature [31], we made the assumption that our topics could be viewed as variations on certain central themes, which were being obscured by the inherent stochastic nature of the LDA technique. In order to determine what these central themes were, we created an aggregate dataset composed of 500 runs of our chosen LDA model, each using the parameters we had defined above (10 topics, multiple $\alpha$ values, and removal of words that appeared less than 15 times or in greater than 55% of the dataset), appending the resulting topics into a matrix. We then ran a clustering analysis on that matrix, using a Gaussian Mixture Modeling technique, which can be thought of as a more advanced variant of clustering techniques like K-means that allows for non-spherically shaped clusters. We set our clustering coefficient equal to the number of topics K produced by each model (10), and extracted the centroids of each cluster. Finally, we used a cosine similarity metric between these centroids and the 10 topics for each model in our dataset to search for the single model which has the greatest degree of overlap with the cluster centers. This “most central model" was then chosen for our final analysis.

Although somewhat abstract, this aggregation technique has a fairly intuitive interpretation, analogous to the data collection approaches taught in introductory physics. Our original observation, that each model produced slightly varying topics, can be thought of as similar to the commonly encountered problem of finding random variance in a set of measurements. In such cases, one technique is to take many measurements, assuming that they will cluster together, and then computing the mean and standard deviation of the resulting cluster. Each of our 500 models provided one set of 10 “measurements” (that is, topics) which could be thought of as points in the high-dimensional “word space” of our data corpus. Our clustering technique allowed us to find the mean of each cluster of measurements. We then chose a single model whose topics fell closest to each of these mean values, which was used for our final analysis of the literature.

We realize that this model selection process calls into question the objectivity of our analysis. However, as previously mentioned, due to its unsupervised nature, objectivity (in the form of producing some kind of “ground truth”) is not the goal of this type of analysis. Rather, our goal is to produce a *replicable*, *validated, independent model* of the data that reveals interesting trends and structures. By replicable, we mean that our methods are transparent and our code can be re-run to produce the same output by any researcher with the requisite computational tools[4]. By independent, we mean that we do not specify any specific desired results or theoretical commitments in advance—although we made various decisions along the line when processing our data, determining model parameters, and selecting our final model, the actual model results arise from an unsupervised algorithm that we do not directly control. By validated, we mean that the process of model selection must be informed by multiple levels of validation, as discussed above. By model, we mean that these results should be evaluated with all of the previously-discussed simplifications and assumptions in mind—that is, as a simplified perspective on an incredibly diverse and complex set of literature.

### D. Analysis of literature using the final model

Once we had identified our most representative model, we used it to perform several analyses on our document corpus to both validate that model and answer our research questions. Our first analysis involved finding the papers that were most aligned with each topic (the “most representative papers”) to provide some face validity for our topics as well as to characterize the meaning of each topic. This type of analysis is a built-in function of a trained LDA model—in fact, it can also be used to evaluate the topic mixtures of new, unseen documents (for example, future PERC proceedings papers), which we see as a promising area of future research. This analysis, along with the topics derived from our model, allowed us to address our first research question, “Based on an LDA analysis of a selection of the Physics Education research literature, what approaches and theories have been prevalent over the last two decades?”

Our second analysis involved graphing the “prevalence” of each topic over time, where prevalence was defined as the aggregated percentage of each topic across all documents in a given year. This leveraged the mixed-membership aspect of LDA by adding up the fractional contributions of each document to the different topics over time. Using this analysis, we can see large-scale shifts in the community over time, which may or may not have been visible in individual papers. In other words, this analysis allows us to see if certain topics have become more or less commonly addressed over time, even if those topics are uniformly distributed across many papers. By aggregating the prevalence of such topics, we can identify broad trends and areas of focus over the last 18 years of PERC proceedings. This analysis helped answer our second research question, "How has the prevalence of these approaches and theories evolved over time?"

[4]A selection of code used for this analysis can be found at https://github.com/uio-ccse/PERC_TopicModel

## IV. RESULTS

### A. Topic Characterizations and Most Representative Papers

We begin by presenting the topics produced by our model, in order of their appearance within the model, shown in Table 1. For each topic, we provide the name we have assigned and the 10 most highly-weighted words in that topic, both alone and with their relative weights. As a reminder, each of these topics is actually a probabilistic distribution over the 4645 unique words and bi-grams in our corpus (our "bag of words"), with different weights assigned to each word. The weights on words are the normalized probability of choosing a word, given that specific topic—so, for example, the weight of 0.014 on the word "representation" in the first topic means that, on average, there would be a 1.4% chance of randomly choosing this word from the bag of words in our model each time one writes about the "Representations" topic.

To clarify the meaning of each topic, in Table 2 we also present the 5 most representative papers according to the model, including the year of publication, title of the paper, authors, and the model's rating of that topic's prevalence within the paper. This rating of prevalence within the paper can be thought of as a kind of "percent match" between the paper and topic—a value of 0.9 means that the model rated the paper in question as being 90% composed of that topic, with the remaining 10% being a mixture across all other topics.

We then explicitly discuss each of the topics in greater detail and clarify our chosen labels for each topic. Thereafter, in section V, we discuss the validity of these groups and potential limitations of this analysis technique.

| Topic Number | Topic Name | Top 10 Words | Top 10 Words with Weights |
|---|---|---|---|
| 1 | Representations | representation, diagram, force, equation, vector, charge, solve, difficulty, mathematical, energy | 0.014*"representation"<br>0.013*"diagram"<br>0.011*"force"<br>0.009*"equation"<br>0.009*"vector"<br>0.008*"charge"<br>0.008*"solve"<br>0.007*"difficulty"<br>0.007*"mathematical"<br>0.007*"energy" |
| **2** | Problem-Solving | problem_solve, solution, task, solve, write, strategy, expert, rubric, principle, feedback | 0.021*"problem_solve<br>0.019*"solution"<br>0.014*"task"<br>0.013*"solve"<br>0.011*"write"<br>0.008*"strategy"<br>0.008*"expert"<br>0.007*"rubric"<br>0.007*"principle"<br>0.007*"feedback" |
| **3** | Labs | lab, activity, experiment, laboratory, simulation, experimental, circuit, report, section, code | 0.042*"lab"<br>0.028*"activity"<br>0.022*"experiment"<br>0.012*"laboratory"<br>0.010*"simulation"<br>0.010*"experimental"<br>0.009*"circuit"<br>0.007*"report"<br>0.006*"section"<br>0.006*"code" |
| **4** | Quantitative Assessment of Concepts | item, correct, score, assessment, post_test, version, choice, force, pre_test, instrument | 0.031*"item"<br>0.011*"correct"<br>0.011*"score"<br>0.010*"assessment"<br>0.010*"post_test"<br>0.009*"version"<br>0.009*"choice"<br>0.007*"force"<br>0.007*"pre_test"<br>0.007*"instrument" |
| **5** | K-12 | teacher, classroom, activity, content, school, curriculum, scientific, practice, high_school, program | 0.052*"teacher"<br>0.014*"classroom"<br>0.011*"activity"<br>0.010*"content"<br>0.009*"school"<br>0.009*"curriculum"<br>0.009*"scientific"<br>0.008*"practice"<br>0.008*"high_school"<br>0.007*"program" |
| **6** | Difficulties with Quantum Mechanics | state, vector, difficulty, quantum_mechanic, quantum, interview, measurement, particle, write, qm | 0.026*"state"<br>0.017*"vector"<br>0.012*"difficulty"<br>0.012*"quantum_mechanic"<br>0.011*"quantum" |
| | | | 0.010*"interview"<br>0.009*"measurement"<br>0.009*"particle"<br>0.008*"write"<br>0.007*"qm" |
| **7** | Community, Identity | community, program, participant, identity, practice, project, interview, faculty, stem, graduate | '0.016*"community"<br>0.013*"program"<br>0.011*"participant"<br>0.010*"identity"<br>0.009*"practice"<br>0.008*"project"<br>0.007*"interview"<br>0.006*"faculty"<br>0.006*"stem"<br>0.006*"graduate" |
| **8** | Qualitative Methodologies and Constructivist Theory Building | energy, resource, interview, object, explanation, say, go, move, talk, frame | 0.020*"energy"<br>0.011*"resource"<br>0.011*"interview"<br>0.007*"object"<br>0.006*"explanation"<br>0.006*"say"<br>0.006*"go"<br>0.005*"move"<br>0.005*"talk"<br>0.005*"frame" |
| **9** | Research-Based Instruction | tutorial, faculty, la, ta, lecture, exam, semester, section, classroom, practice | 0.022*"tutorial"<br>0.021*"faculty"<br>0.021*"la"<br>0.018*"ta"<br>0.018*"lecture"<br>0.011*"exam"<br>0.011*"semester"<br>0.008*"section"<br>0.007*"classroom"<br>0.007*"practice" |
| **10** | Quantitative Surveys of Demographic Gaps | score, performance, survey, average, grade, gain, measure, gender, factor, semester | 0.040*"score"<br>0.017*"performance"<br>0.014*"survey"<br>0.013*"average"<br>0.011*"grade"<br>0.011*"gain"<br>0.011*"measure"<br>0.008*"gender"<br>0.008*"factor"<br>0.008*"semester" |

**Table 1:** Topic numbers, names, top 10 words, and top words with weights

| Topic | Year | Title | Authors | Prevalence in Paper |
|---|---|---|---|---|
| Representations | 2017 | Student determination of differential area elements in upper-division physics | Benjamin P. Schermerhorn and John R. Thompson | 0.90 |
| | 2012 | Arrows as anchors: Conceptual blending and student use of electric field vector arrows | Elizabeth Gire and Edward Price | 0.88 |
| | 2016 | The impact of students' epistemological framing on a task requiring representational consistency | Alexandru Maries, Shih-Yin Lin, and Chandralekha Singh | 0.87 |
| | 2011 | Student Interpretation of the Signs of Definite Integrals Using Graphical Representations | Rabindra R. Bajracharya, Thomas M. Wemyss, and John R. Thompson | 0.87 |
| | 2012 | Contrasting students' understanding of electric field and electric force | Alejandro Garza and Genaro Zavala | 0.83 |
| Problem-Solving | 2012 | Transfer of Argumentation Skills In Conceptual Physics Problem Solving | Carina M. Rebello and N. Sanjay Rebello | 0.91 |
| | 2015 | How do Multimedia Hints Affect Students' Eye Movements in Conceptual Physics Problems? | Xian Wu, John Hutson, Lester C. Loschky, and N. Sanjay Rebello | 0.87 |
| | 2004 | Computer Problem-Solving Coaches | Leonardo Hsu and Kenneth Heller | 0.82 |
| | 2015 | Effects of Visual Cues and Video Solutions on Conceptual Tasks | Tianlong Zu, Elise Agra, John Hutson, Lester C. Loschky, and N. Sanjay Rebello | 0.80 |
| | 2003 | Expert-Novice Comparisons to Illuminate Differences in Perceptions of Problem Solutions | Kathleen A. Harper | 0.79 |
| Lab Instruction | 2007 | Using Students' Design Tasks to Develop Scientific Abilities | Xueli Zou | 0.81 |
| | 2004 | Can Computer Simulations Replace Real Equipment in Undergraduate Laboratories? | Noah D. Finkelstein, Katherine K. Perkins, Wendy K. Adams, Patrick B. Kohl , and Noah S. Podolefsky | 0.79 |
| | 2006 | When and How Do Students Engage in Sense-Making in a Physics Lab | Anna Karelina and Eugenia Etkina | 0.76 |
| | 2017 | Using lesson design to change student approaches to dorm-room design prelabs | Katherine Ansell and Mats Selen | 0.76 |
| | 2016 | Using RealTime Physics with different instructional technologies in a circuits lab | Mónica Quezada-Espinoza, Angeles Dominguez, and Genaro Zavala | 0.74 |
| **Quantitative Assessment of Concepts** | 2016 | Assessing Student Learning and Improving Instruction with Transition Matrices | Paul J. Walter and Gary A. Morris | 0.95 |
| | 2017 | Showing the dynamics of student thinking as measured by the FMCE | Trevor I. Smith, Kerry A. Gray, Kyle J. Louis, Bartholomew J. Ricci, and Nicholas J. Wright | 0.93 |
| | 2018 | Determining a hierarchy of correctness through student transitions on the FMCE | Kyle J. Louis, Bartholomew J. Ricci, and Trevor I. Smith | 0.93 |
| | 2012 | Exploring student difficulties with pressure in a fluid | Matthew Goszewski, Adam Moyer, Zachary Bazan, and Doris J. Wagner | 0.93 |
| | 2016 | A Multi-faceted Approach to Measuring Student Understanding | Ian T. Griffin, Kyle J. Louis, Ryan Moyer, Nicholas J. Wright, and Trevor I. Smith | 0.89 |
| **K-12** | 2007 | How Elementary Teachers Use What We Teach: The Impact of PER At The K-5 Level | Danielle B. Harlow | 0.94 |
| | 2011 | Teacher-driven professional development and the pursuit of a sophisticated understanding of inquiry | Michael J. Ross, Ben Van Dusen, Samson Sherman, and Valerie K. Otero | 0.89 |
| | 2005 | Different Views on Inquiry: A Survey of Science and Mathematics Methods Instructors | Thomas Withee and Rebecca S. Lindell | 0.87 |
| | 2010 | Uncovering the Hidden Decisions that Shape Curricula | Danielle B. Harlow | 0.87 |
| | 2010 | Learning Pedagogy in Physics | Danielle B. Harlow, Lauren Swanson, Hilary A. Dwyer, and Julie A. Bianchini | 0.86 |
| **Difficulties with Quantum Mechanics** | 2011 | Students' Difficulties with Quantum Measurement | Guangtian Zhu and Chandralekha Singh | 0.97 |
| | 2015 | Student difficulties with quantum states while translating state vectors in Dirac notation to wave functions in position and momentum representations | Emily Marshman and Chandralekha Singh | 0.96 |
| | 2018 | Student difficulties with the corrections to the energy spectrum of the hydrogen atom for the intermediate field Zeeman effect | Emily Marshman, Christof Keebaugh, and Chandralekha Singh | 0.96 |
| | 2006 | Student Difficulties with Quantum Mechanics Formalism | Chandralekha Singh | 0.95 |
| | 2014 | Developing an Interactive Tutorial on a Quantum Eraser | Emily Marshman and Chandralekha Singh | 0.95 |
| **Community and Identity** | 2016 | Improving representation in physical sciences using a Departmental Action Team | Katherine Rainey, Joel C. Corbo, Daniel L. Reinholz, and Meredith Betterton | 0.96 |
| | 2018 | Characterizing Models of Informal Physics Programs | Claudia Fracchiolla, Noah D. Finkelstein, and Kathleen A. Hinko | 0.93 |
| | 2018 | The Intersection of Identity and Performing Arts of Black Physicists | Tamia Williams, Simone Hyater-Adams, Kathleen A. Hinko, Claudia | 0.92 |
| | | | Fracchiolla, Kerstin Nordstrom, and Noah D. Finkelstein | |
| | 2017 | An Analysis of Community Formation in Faculty Online Learning Communities | Alexandra Lau, Melissa H. Dancy, Joel C. Corbo, Charles R. Henderson, and Andy Rundquist | 0.92 |
| | 2018 | Intense Outreach: Experiences Shifting University Students' Identities | Brean Prefontaine, Claudia Fracchiolla, Manuel Vasquez, and Kathleen A. Hinko | 0.91 |
| **Qualitative Methodologies and Constructivist Theory Building** | 2018 | Examining the productiveness of student resources in a problem-solving interview | Lisa M. Goodhew, Amy D. Robertson, Paula R. L. Heron, and Rachel E. Scherr | 0.94 |
| | 2010 | Generating Explanations for an Emergent Process: The Movement of Sand Dunes | Lauren Barth-Cohen | 0.93 |
| | 2012 | Examining the Positioning of Ideas in the Disciplines | Vashti Sawtelle, Tiffany-Rose Sikorski, Chandra Turpen, and Edward F. Redish | 0.91 |
| | 2007 | Conceptual Dynamics in Clinical Interviews | Bruce L. Sherin, Victor R. Lee, and Moshe Krakowski | 0.89 |
| | 2011 | Intuitive ontologies for energy in physics | Rachel E. Scherr, Hunter G. Close, and Sarah B. McKagan | 0.87 |
| **Research-Based Instructional Methods** | 2009 | Towards Understanding Classroom Culture: Students' Perceptions of Tutorials | Chandra Turpen, Noah D. Finkelstein, and Steven J. Pollock | 0.94 |
| | 2009 | Student Perspectives on Using Clickers in Upper-division Physics Courses | Katherine K. Perkins and Chandra Turpen | 0.89 |
| | 2011 | But Does It Last? Sustaining a Research-Based Curriculum in Upper-Division Electricity & Magnetism | Stephanie V. Chasteen, Rachel E. Pepper, Steven J. Pollock, and Katherine K. Perkins | 0.86 |
| | 2007 | Research-based Practices For Effective Clicker Use | C. J. Keller, Noah D. Finkelstein, Katherine K. Perkins, Steven J. Pollock, Chandra Turpen, and Michael Dubson | 0.86 |
| | 2009 | The Impact of Physics Education Research on the Teaching of Introductory Quantitative Physics | Charles R. Henderson and Melissa H. Dancy | 0.83 |
| **Quantitative Assessment of Demographic Gaps** | 2017 | Racial and ethnic bias in the Force Concept Inventory | Rachel Henderson and John Stewart | 0.95 |
| | 2008 | The Persistence of the Gender Gap in Introductory Physics | Lauren E. Kost, Steven J. Pollock, and Noah D. Finkelstein | 0.94 |
| | 2014 | The Impacts of Instructor and Student Gender on Student Performance in Introductory Modeling Instruction Courses | Daryl McPadden and Eric Brewe | 0.92 |
| | 2010 | Improved Student Performance in Electricity and Magnetism Following Prior MAPS Instruction in Mechanics | Saif Rayyan, Andrew Pawl, Analia Barrantes, Raluca E. Teodorescu, and David E. Pritchard | 0.92 |
| | 2010 | Gender Differences in Physics 1: The Impact of a Self-Affirmation Intervention | Lauren E. Kost-Smith, Steven J. Pollock, Noah D. Finkelstein, Geoffrey L. Cohen, Tiffany A. Ito, and Akira Miyake | 0.92 |

**Table 2:** Topics, along with the 5 most representative papers, including title, authors, and percent match with that topic according to the LDA model

**Topic 1: Use and understanding of representations.**
This topic focuses on research on student understanding of and difficulties with representations, including mathematical representations (words: "equation", "mathematical") and graphical representations ("diagram, vector") as well as some specific physics topics frequently used in this strand of research ("charge", "force"). The most representative papers for this topic are all about student use and understanding of particular representational forms like graphs, vectors, and integrals, consistency across these representational forms, or subjects like electricity and magnetism that require students to become fluent in different representations.

**Topic 2: Problem-solving strategies and instructor support.**
This topic focuses on research on problem solving in physics, including words related to problem-solving expertise ("expert", "strategy") and feedback ("rubric", "feedback"). Representative papers for this topic focus on different facets of problem-solving such as argumentation skills, computer aids, video solutions, and expert/novice differences.

**Topic 3: Laboratory instruction theory and practice.**
This topic focuses on research on physics laboratories, including the written components ("report") and specific activities ("simulation", "circuit"). The word "code" here might refer to computer code, or the process of coding data for analysis. The most representative papers for this topic focus on different approaches to laboratory instruction such as design tasks, computer simulations, and active learning transformations.

**Topic 4: Quantitative assessment of conceptual understanding.**
This topic focuses on use of quantitative methods to measure student conceptual understanding, with a specific focus on concept inventories. The topic includes a mix of quantitative assessment-related words (such as "item", "correct", "score", "pre_test", "post_test", "choice") and concept-inventory words (such as "force", "instrument", and "version"). Representative papers for this topic focus on assessment of student understanding, application of existing concept inventories, methods for analyzing concept inventory results, and development of new concept inventories.

**Topic 5: K-12 teacher training, professional development, and curriculum design.**
This topic focuses on research physics teaching below the college level, including teacher professional development and pedagogy. It includes a number of words that are usually used to refer to teaching in K-12 settings ("teacher", "high_school", "school", "teacher") as well as several more general teaching-related words that could encompass multiple contexts ("content", "curriculum", "scientific", "practice", "program"). The most representative papers for this topic focus primarily on K-12 teacher professional development, with some papers also focusing on K-12 curriculum design.

**Topic 6: Student understanding of and difficulties with quantum mechanics.**
This topic focuses on student understanding of and difficulties with quantum mechanics. It includes numerous quantum mechanics-related words, as well as two words that are likely more methodological in nature ("interview", "write"). Representative papers for this topic focus exclusively on student difficulties with various aspects of quantum mechanics such as the concepts of measurement and energy levels, notation, as well as development of tutorials to help with these difficulties.

**Topic 7: Communities of practice, identities, and institutional change.** This topic focuses on research on student identities and physics education communities, with a specific focus on institutional change. It includes words such as "community" and "practice", which we suspect might refer to the communities of practice framework, and the word "identity." It also includes words that likely refer to institutional change ("program", "faculty", "stem", "graduate"). Representative papers for this topic focus on the development of different types of communities, both for physics learning and for departmental change, and intersectional analyses of physics student identities.

**Topic 8: Qualitative methodologies and constructivist theory building.** This topic focuses on more qualitative, theoretical case studies of physics student cognition with a specific focus on student discourse and explanation. The topic includes several words commonly associated with studies of student cognition and understanding ("resource", "frame") and references to qualitative methodologies such as interviews and observations ("interview", "explanation", "say", "talk"). We suspect the primary word, "energy", refers to the fact that many of the papers aligned with this topic used the concept of energy as the foci for their case studies (e.g., [32]). Representative papers for this topic focus on interview methodologies, student explanations, and cognitive theories of student understanding such as conceptual resources or ontological categories.

**Topic 9: Use of research-based instructional techniques.** This topic focuses on use of research-based instructional techniques, such as tutorials and clickers, in lectures and recitation sections. It includes words related to specific techniques ("tutorial"), settings ("lecture", "classroom", "section"), assessment ("exam"), and implementation ("faculty", "la", "ta"). The most representative papers for this topic focus on use of research-based instructional tools and the effects of these instructional tools on student learning.

**Topic 10: Quantitative assessment of demographics-based learning gaps.** This topic focuses on quantitative PER research about topics other than conceptual understanding, with an especial focus on gender and ethnicity. It includes one overlapping word with Topic 4 ("score") and several other words related to quantitative measurement and analysis ("performance", "survey", "average", "grade", "gain", "gender", "factor"). Representative papers for this topic mostly focus on measuring the effects of gender and race/ethnicity on student concept inventory performance, with a smaller subset focusing on assessing the impact of modeling instruction interventions.

## B. Topic prevalence over time

In this section, we examine the changing prevalence of topics over time. We define "prevalence" as the sum of particular topic percentages across documents in a particular year, which can be measured both cumulatively and averaged by year. For example, if in a certain year all documents were evaluated by the model to contain 10% of Topic 1, the average prevalence of Topic 1 for that year would be 10%, while the cumulative prevalence would be 0.1 x N, where N is the number of documents produced that year. Summing the cumulative prevalence across all topics for a particular year would return the total number of documents in that year. Because LDA is a mixed-membership model, evaluating topic prevalence in this way allows us to see large-scale trends across the PER community that might be thinly distributed across many documents.

Plots of cumulative prevalence over time for each of the 10 topics are shown in Figure 6. Based on our definition of cumulative prevalence, the y-axis of the figure can be thought of as a count of the "effective number of papers" produced on a topic each year. For example, a cumulative prevalence of 5.0 for the Representations topic would indicate that the equivalent of 5 complete papers about student understanding of representations was produced that year. We use the term "effective" because in practice every topic is distributed (to a varying degree) across the entire corpus.

In order to get a sense for how distributed these topics are each year, we have used a repeated 5-fold cross validation method[5] to estimate the variance of prevalence values. Under this method, for each topic-year we randomly split the cumulative prevalence values into 5 equal "chunks," each of which contains 20% of the values for that year. We then re-calculate the cumulative prevalence 5 times, each time leaving out one of the chunks for that year (and normalizing for this missing 20%). We then repeat this process 20 times, randomly re-splitting the data for each new iteration. We then take the standard deviation, $\sigma$, of the resulting set of measurements, and use $3\sigma$ as a measure of the variance. Based on this approach, the error bars can be thought of as a measure of how concentrated each topic is into a small number of papers in a particular year, with their size being proportional to this level of heterogeneity. That is, if a topic is evenly distributed across all topics in one year, the 5-fold cross validation method described above will produce a spread of 0. If, on the other hand, a topic is strongly concentrated in only a few documents, the samples that "leave out" those documents will lead to significantly varying prevalence measurements, leading to a larger spread.

[5] Necessary because prevalence counts for each topic/year are not Gaussian distributed

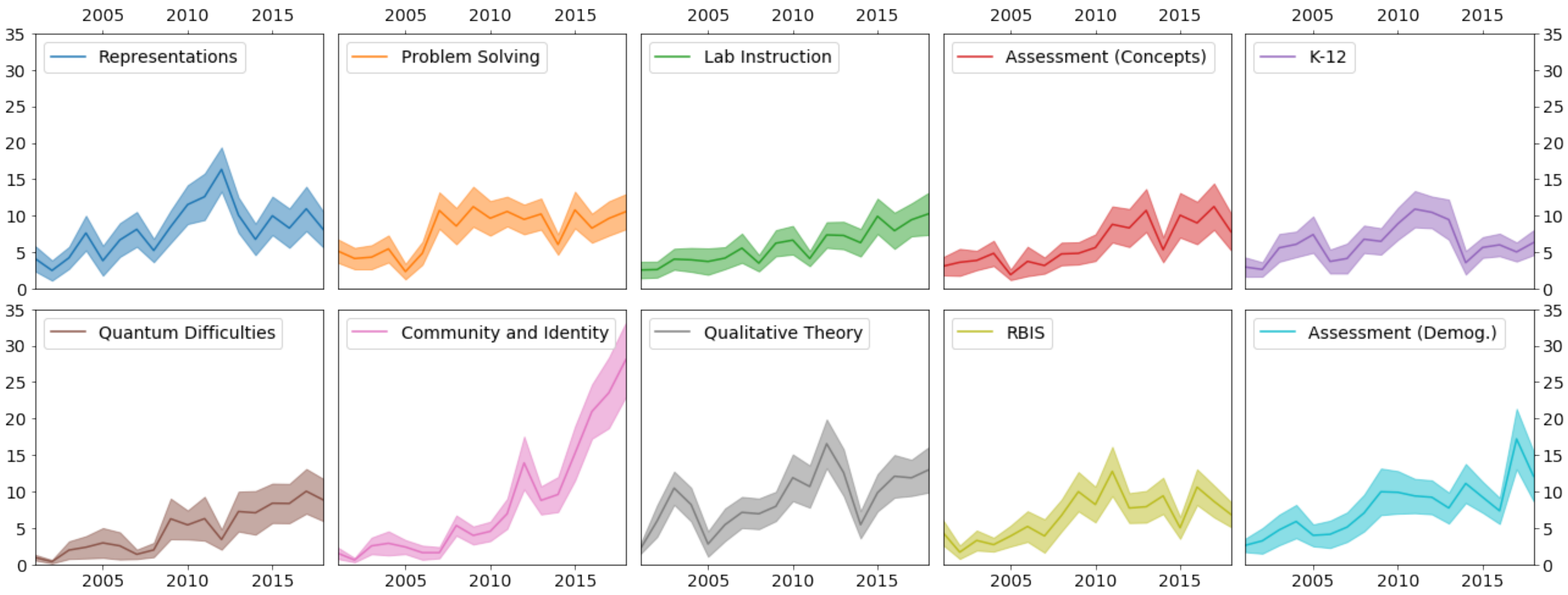

**Figure 6:** Cumulative topic prevalence over time, with error bars based on 5-fold cross validation. Wider error bars correspond to a greater concentration of topics in a small number of papers, while narrower error bars indicate a more uniform distribution. Y-axis corresponds to the effective number of papers each year on that topic.

The results shown in Figure 6 suggest that all of the topics have seen some sustained growth over time. This, however, is not surprising as there are significantly more papers being published in the PERC proceedings in recent years than in the early years. Certain topics have small spikes at different points in the history of PERC: specifically, the Representations topic has a spike around 2013; Problem Solving has a dramatic uptick after 2005; and the Qualitative Theory-Building topic has spikes around 2003 and 2012. Both the Qualitative Theory-Building topic and the K-12 topic also see significant downward shifts around 2012-2013. By far the most dramatic visible trend is in the "Communities of practice, identities, and institutional change" topic (Topic 7) which has seen an especially sharp upward trend since 2010, to nearly 30 effective papers per year in recent years. The width of the error bars on this trend (proportional to the level of concentration of the topic within a set of papers in a certain year) indicates that this topic seems to be more concentrated within a small number papers, suggesting that this trend may be due to the concerted efforts of a dedicated group of researchers rather than an overall shift within the community.

In order to reveal trends over time, we have used a data-smoothing technique which dampens out some of the minor year-to-year variations. Specifically, we have used a "rolling window" function of width 3, which averages the prevalence values for each year with those of the preceding and following year. Although this technique neglects the values of the first and last years, it serves dampen out small shifts from year to year while keeping larger-scale trends intact allowing us to more clearly see how these trends have developed over time. These smoothed graphs of cumulative and average prevalence over time are shown in figures 7 and 8.

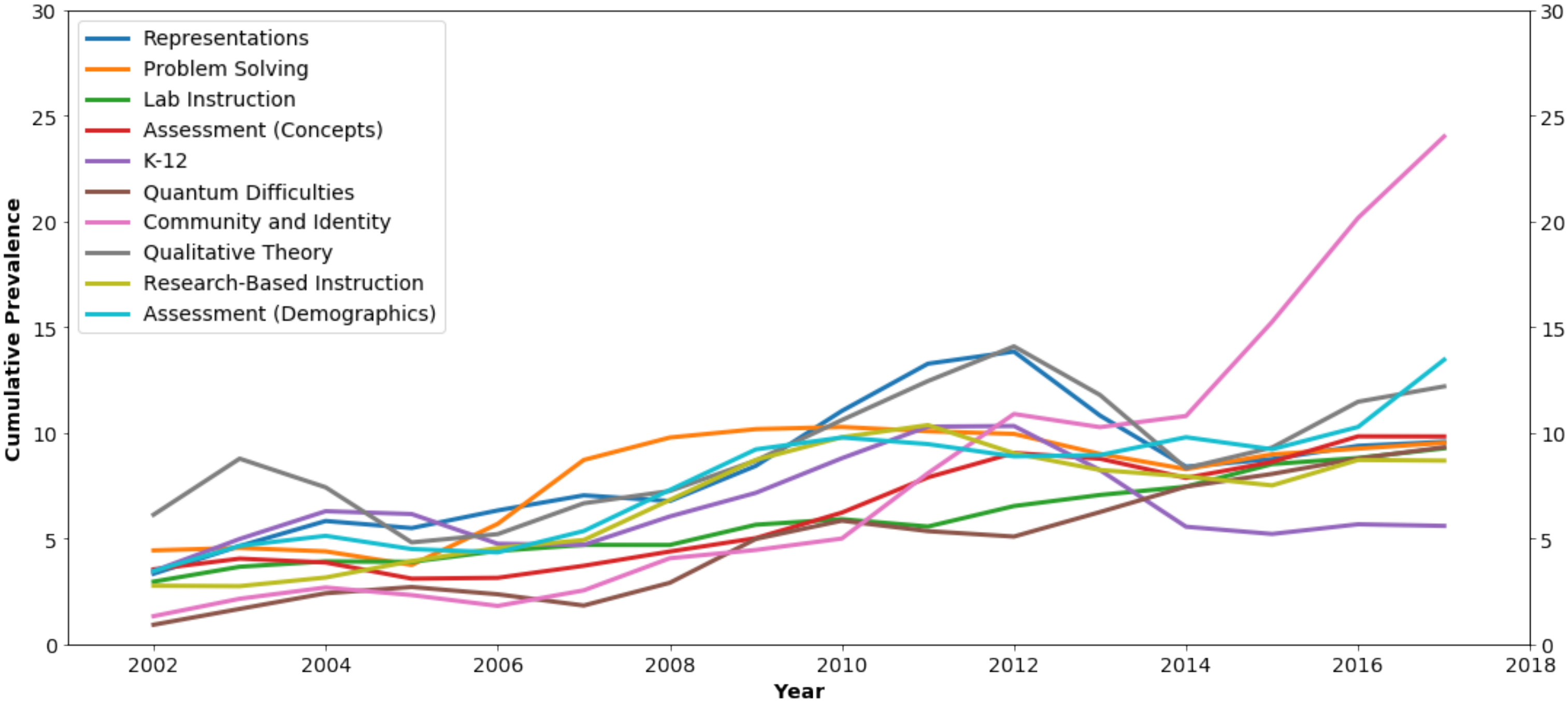

**Figure 7:** Cumulative prevalence of topics over time, smoothed using a rolling 3-year window

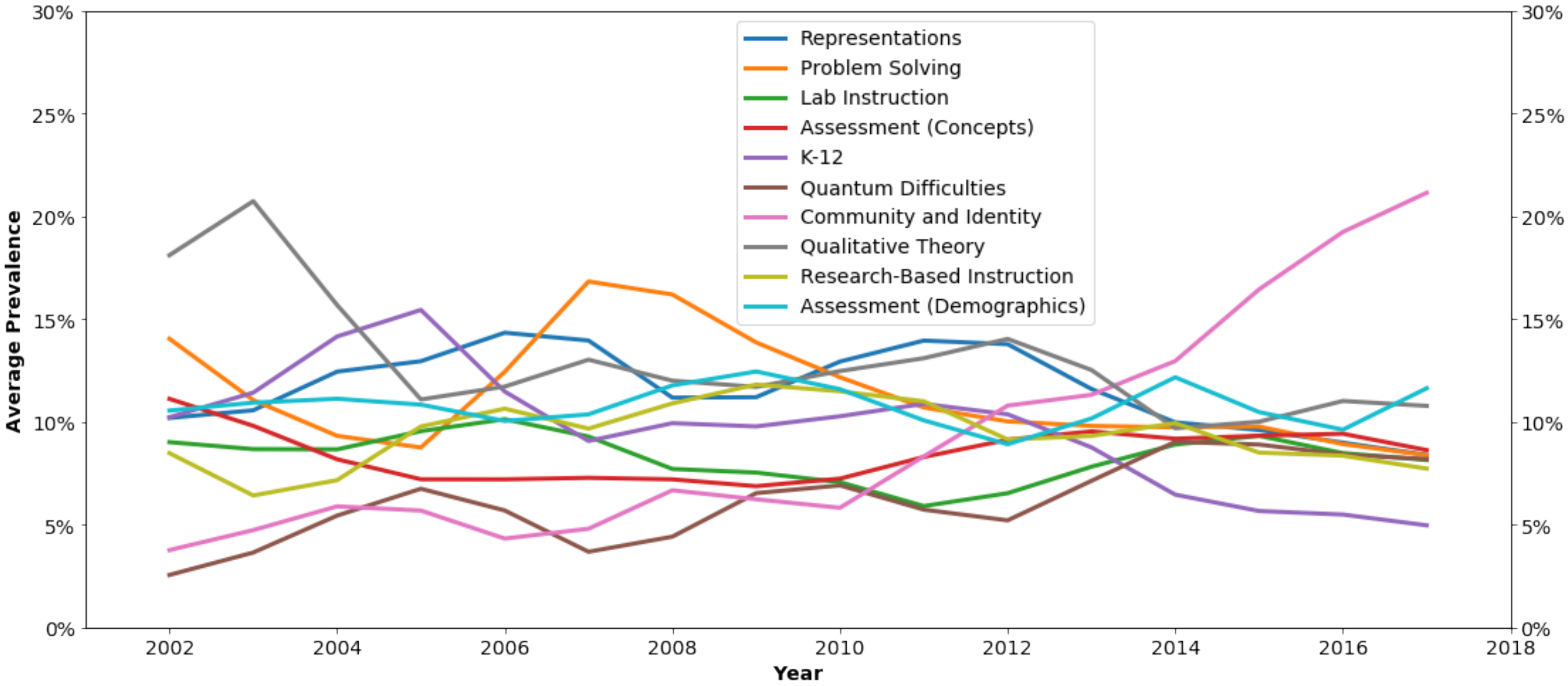

**Figure 8:** Average prevalence of topics over time, smoothed using a rolling 3-year window

Because the number of documents published in PERC proceedings has increased substantially over its years of publication, we present both cumulative and average prevalence over time in Figures 7 and 8. The graph of cumulative prevalence in Figure 7 allows us to see the magnitude of changes throughout the years, while the average prevalence graph in Figure 8 allows us to see shifts over time that are normalized for the growing PER membership.

Looking at these graphs, we first notice that many of the topics have remained relatively constant over time, even though they are not necessarily very widespread overall. For example, the Quantum Difficulties topic remains fairly consistent on Figure 8, and roughly tracks with the other topics based on cumulative prevalence. We suspect that this consistency is due to the fact that, unlike other topics which incorporate multiple research methods and subjects, this topic has focused on a specific subject, based on a distinct vocabulary, over a long period of time. Similarly, both assessment topics remain fairly stable over time on average, which might indicate that this method has seen continued use in recent years through the regular development of targeted concept inventories (e.g., [33,34]).

Beyond this overall stability, there are three trends immediately visible on both graphs: First, there is a spike in topic 8, "Qualitative methodologies and constructivist theory building" during the early years of PER. We suspect this spike may be related to the early push to construct and adapt cognitivist learning theories to physics education (e.g., [35,36]). A smaller spike in this topic, along with a parallel spike in topic 1 ("Representations") occurs around 2012. Second, there is a wave of research on problem-solving starting in about 2006, peaking in 2008, and decreasing from there (relative to the rest of the topics). Although we cannot say for certain where this trend comes from, based on the representative papers for this topic we suspect this trend corresponds to a simultaneous focus on problem solving by several PER groups in the Midwestern US, especially Kansas State University (e.g., [37,38]). Third, we again see a major increase in the "Communities of practice, identities, and institutional change" topic prevalence since about 2011. This is by and far the largest trend on the graph, and based on its low prevalence prior to that point, this seems to be a new trend in the field. Although the most recent year is absent in the smoothed graphs of prevalence over time, one can see from Figure 6 that the upward trend has continued through 2018. Based on the combined focus on identities, communities of practice, and institutional change visible in the topic's words and representative papers, we would argue that this trend indicates an increasing shift in the way we, as a community, are conceptualizing teaching and learning physics, towards a more sociocultural and community-based perspective.

## V. DISCUSSION: IMPLICATIONS AND LIMITATIONS OF LDA MODEL RESULTS

To answer our first research question, using LDA we have identified 10 enduring themes in PER that have been explored over the last two decades: 1) Research on representations; 2) Research on problem-solving; 3) Research on laboratory instruction, 4) Assessment of student understanding; 5) Research on physics teacher preparation and physics instruction at the K-12 level; 6) Research on student difficulties with quantum mechanics; 7) Research on communities of practice, identities, and institutional change; 8) Qualitative, constructivist theory building; 9) Research-based instructional methods; 10) Assessment of demographics-based learning gaps.

To answer our second research question, this analysis provides suggestions for where the field has come from, and where it seems to be going. Based on the trends discussed above, it appears that much of the research in PER has remained fairly steady over the years and that there remains a strong, enduring interest in all of the research topics mentioned above. However, the field has also had waves of increased interest in several of the topics, beginning with initial interest in qualitative, theory-building studies of student cognition. Over time, that gave rise to an increased focus on problem-solving. In more recent years, the field has increasingly moved towards more sociocultural views of teaching and learning, with an especial focus on communities of practice, student identities, and institutional change. We note that certain researchers (ourselves included) have anecdotally noticed an increase in studies leveraging these types of sociocultural frameworks in recent years, and this model provides support for that observation.

Because this model is meant to provide an overview of the research being done within the PER community as a whole, it seems useful to contrast it with other such work, for example the 2014 PER literature review compiled by Docktor and Mestre [3]. In that review, they break the field apart into six primary topical areas of study:

- Conceptual understanding
- Problem solving
- Curriculum and instruction
- Assessment
- Cognitive psychology
- Attitudes and beliefs about teaching and learning

Comparing our model to this review, there are several immediate points of overlap: for example, both our model and Docktor and Mestre's review include physics problem-solving by name. Both also contain themes explicitly focused on assessment, though the research on assessment described by Docktor and Mestre seems to encompass both of our assessment-related topics, Topics 4 (Assessment: Concepts) and 10 (Assessment: Demographics). Docktor and Mestre's Curriculum and Instruction topical area is defined to cover research on lectures, recitations, and laboratories, which solidly encompasses our Topic 3 (Lab Instruction) and Topic 9 (Research-Based Instructional Strategies). We also suspect that their Conceptual understanding topic would likely cover our Topic 6 (Quantum difficulties) and Topic 8 (Qualitative Theory-Building). All in all, our model seems to have significant overlap with the topical areas they define.

Of course, there are also some areas that do not overlap. For starters, our model has grouped together research on representations (and to some degree electricity and magnetism concepts) into its own topic, while Docktor and Mestre place research on representations under the heading of problem solving. Docktor and Mestre deliberately excluded research on pre-college physics education and physics teacher preparation, while our model explicitly covers this research under the heading of Topic 5 (K-12). We also do not see any reference to research on student identities, communities of practice, or institutional change in their review, although we note that their article was published in 2014 which (if the trends discussed above are to be believed) was before research on this topic had really begun to take off. On the other hand, our model does not explicitly recognize research on either cognitive psychology or attitudes and beliefs about teaching and learning as their own distinct topics.

In making these comparisons, we do not mean to question the usefulness or thoroughness of large-scale reviews like Docktor and Mestre's. Rather, we simply want to use this as an argument for the fact that our model has produced results that seem to be well in line with many longstanding and long-recognized trends in the field. Thus, as previously mentioned, we would argue that this model provides a quantitative, independent, and replicable (but not objective) approach to analyzing the field.

Beyond these results, we would also argue that this study serves a methodological purpose: it provides a proof-of-concept that natural language processing techniques can successfully be used to analyze large amounts of educational research literature. However, we wish to be clear about the limitations of this type of analysis. One is limitation with this method is that it is particularly sensitive to research on distinctive topics that is sustained over time. This type of sensitivity is inherent to the way LDA classification works—if there are distinctive sets of words that consistently co-occur, they are likely to be grouped into a single topic. We suspect this is part of the reason why research on student difficulties with quantum mechanics was identified as its own topic, despite the fact that it is not very widespread throughout the community—it is a topic of study that has been sustained over time, which uses distinct terminology uncommon to the remainder of the literature. LDA will also tend to associate texts that use a similar set of vocabulary, even if they are referring to different phenomena; for example, the Community and Identity topic includes research on both informal learning communities and institutional change communities. Although a shared vocabulary might indicate a shared set of theoretical commitments (for example, sociocultural views of teaching and learning) the set of work encompassed by these kinds of topics is significantly more diverse than, for example, the work grouped into the Laboratory Instruction topic.

Additionally, returning to our original description of the LDA method, we wish to revisit to the many simplifying assumptions inherent to this type of topic modeling. Writing a scientific paper is a much more complicated affair than the simplified process assumed by LDA. Additionally, in order to prepare our dataset for analysis, we removed a significant number of words (including some that are frequently used by certain PER communities), strip out much of the semantic meaning of sentences, and remove all numbers and symbols, many of which might be of great importance to a paper's argument. Doing so allows us to analyze large amounts of text in a short amount of time, but this means that the analysis may miss fine-grained distinctions between individual papers. To some degree, this is assumed to be offset by the aggregate trends in word usage (that is, if two papers are actually substantially similar, they will use a similar subset of words), however there is a limit to the amount of information that can be inferred by such trends. As previously stated, this analysis should be considered a kind of "toy model" of the PER literature, and any results it produces should be evaluated with this in mind.

In the course of this project, we also had to grapple with several problems that are not well addressed in the literature, especially the instability of topic models with regard to random initializations. We addressed this limitation using our ensemble clustering method. As a check on this method, in addition to the model presented above we performed two additional 500-model replication runs with different random seed values. Our first replication run featured an extremely similar set of 10 topics (with minor reshuffling of words); the second differed from our presented model by one topic (it had two topics focused on representations and only one on quantitative methods). However, both replication runs showed similar large-scale trends over time to those presented above. This, we feel, provides evidence that our clustering method adequately addressed the instability limitation of LDA.

Finally, we note that many of the model-building decisions made along the way, from the pre-processing steps [24], to the number of models included in the clustering analysis, to the technique chosen for clustering, likely affected our final results. As previously stated, this is in line with standard practice in the field of natural language processing. As Grimmer and Stewart [23] argue, text analysis can be used to gain insights into data, but cannot replace careful analysis by humans. This means that this model should not be taken as a completely "true" result—rather, it is a simplified model of the system under study, with all of the limitations that entails.

Despite all of these limitations, we would like to reiterate the fact that all of these topics and trends, many of which we suspect will be familiar to members of the PER community, arose from an unsupervised machine learning model. Although we did make numerous decisions throughout the modeling process, we did not specify any of the topics a priori—they all arose from the algorithm utilized by the model.

## VI. CONCLUSIONS

In this article, we have presented a method for applying natural language processing techniques to extract the topics of a large educational research literature base. Using this technique, we have been able to define and analyze themes present in the PER literature over the last 18 years, as seen through the PERC proceedings. We hope this analysis will shed some additional light on where the field has come from and where it seems to going.

We see great promise for machine learning techniques like LDA and believe that this kind of work could provide a rich avenue for future research. For example, by analyzing analogous publications to PERC from other fields, one might be able to compare the key topics of physics education research with those from other fields of discipline-based educational research like chemistry, biology, or math education. We also see great promise for analyzing larger data sets, over longer timescales. For example, some journals in the field of science education have over 100 years' worth of publications in their records. These would be very interesting to analyze, especially with an eye to how the practice of physics education has developed over the last century [39].

We also see room for methodological development, for example in exploring the effects of different pre-processing decisions or clustering approaches on model results. These could, for example, reveal other trends in the PER literature not explored here. As previously mentioned, there are extensions to LDA that have been developed in order to analyze additional covariate factors, such as word sentiment [23]. Such methods could potentially be used to add in factors like citation rate to see how they relate to the different topics and provide another metric to evaluate topical prevalence. However, we feel these covariate factors would likely be more useful with full journal articles, such as PRPER or Science Education.

In summary, we look forward to using this method for future explorations of the educational research literature.

## ACKNOWLEDGMENTS

This work was funded by Norges Forskningsrådet Project Number 288125 "International Partnership for Computing in Science Education" and by the Norwegian Agency for Quality Assurance in Education (NOKUT) which supports the Center for Computing in Science Education. We also thank Anders Malthe-Sørenssen, Christine Lindstrøm, Chandra Turpen, Lin Ding, Elise Lockwood, Marieke Kuijjer, and John M. Aiken for their support and feedback on this project.

---

## Appendix A: Mathematical Explanation of LDA's Probabilistic Topic Modeling Technique

LDA is based on a particular probabilistic model for how text documents are generated, which helps to explain the parameters it uses. In this idealized writing process, a corpus, or complete set of documents *D* are "written" through the following generative process [1,19]. First, a writer determines the "topics" they will write about, which are conceptualized as distributions across all possible words that could be included in the document:

1) For each topic $k = \{1,\ldots, K\}$, draw a distribution $\beta_k$ over the set of words, *V,* where $\beta_k \sim \text{Dir}(\eta)$

This determines the how different topics weight each word from the full set of possible words (or vocabulary) V, where the word-topic distribution $\beta_k$ is a multinomial distribution based on the Dirichlet distribution parameterized by $\eta$, Dir($\eta$). The Dirichlet distribution is a multivariable generalization of the Beta distribution, which essentially allows one to specify a set of probable outcomes across (k-1)-dimensional probability simplex. This simplex can be visualized as a kind of "triangle" generalized to k-1 dimensions with k vertices, where each vertex represents 100% of a particular outcome, and locations in between represent mixtures of several outcomes. For example, in the 3-sided distribution (a triangle) shown in Figure 1, the "outcomes" correspond to mixtures across the three different topics. Thus, headlines determined to be entirely aligned with one single topic would be located at the vertices, while those weighted as 50/50 would be on located at the midpoint between two vertices. The Dirichlet parameter $\eta$ specifies how likely points are to fall into a specific corner, with $\eta < 1$ leading to more clumping around the corners, and $\eta > 1$ leading to a smoother distribution across the entire shape.

Once one has determined the topics, the next step is to determine which topics each document will hold:

2) For every document $d \in \{1, \ldots, D\}$ draw a distribution over topics, $\theta_d \sim \text{Dir}(\alpha)$

where the document-topic distribution, $\theta_d$, is also distributed Dirichlet, parameterized by $\alpha$. This means that the particular distribution of topics in the documents is imagined to depend on the Dirichlet parameter α, with small α corresponding to a few topics per document and a large α corresponding to a more even mixture.

Finally, the writer uses these two distributions to generate the words in each document:

3) To generate each of the N words in the document *d*
    a. Choose a topic, $z_{d,n} \sim$ Multinomial($\theta_d$), where $z_{d,n} \in \{1, \ldots, K\}$
    b. Choose a word $w_{d,n} \sim$ Multinomial($\beta_{z_{d,n}}$), where $w_{d,n} \in \{1, \ldots, V\}$

Several of these distributions are latent variables (not known a priori): the set of topics $\beta_K$, the proportion of each topic across documents $\theta_D$, and the assignments of each word to a topic $z_N$. Only the words in each document, $w_N$, can be observed directly. $\alpha$ and $\eta$ are input parameters (commonly known as "hyperparameters") which we specify beforehand. Therefore, the joint distribution of all variables (that is, the probability of generating a particular corpus) becomes:

$$p(\beta_{1:K}, \theta_{1:D}, z_{1:D}, w_{1:D} | \alpha, \eta) = \prod_{k=1}^{K} p(\beta_k | \eta) \prod_{d=1}^{D} p(\theta_d | \alpha) \prod_{n=1}^{N} p\left(z_{d,n} \middle| \theta_d\right) p\left(w_{d,n} \middle| z_{d,n}, \beta_{d,k}\right) \quad (1)$$

Here, $p(\beta_k | \eta)$ is the probability of drawing a particular topic distribution $\beta_k$, given a value of the hyperparameter $\eta$; $p(\theta_d | \alpha)$ is the probability of drawing a certain mixture of topics in a document d, given the hyperparameter $\alpha$; $p\left(z_{d,n} \middle| \theta_d\right)$ is the probability of choosing a specific topic k, given the specific document-topic mixture $\theta_d$; and $p\left(w_{d,n} \middle| z_{d,n}, \beta_{d,k}\right)$ is the probability of choosing a certain word w, given the particular topic k and its corresponding topic-word distribution, $\beta_k$.

In practice, the latent variables must be inferred, since neither $\beta_K$, $\theta_D$, or $z_D$ are known a priori. So, our goal becomes to infer these hidden distributions using statistical inference. That is, we wish to compute the conditional probability of a certain set of topic-word distributions, document-topic distributions, and word-topic assignments, given the observed set of words and Dirichlet parameters:

$$\mathrm{p}(\beta_{1:K}, \theta_{1:D}, z_{1:N} | w_{1:N}, \alpha, \eta) = \frac{p(\beta_{1:K}, \theta_{1:D}, z_{1:D}, w_{1:D} | \alpha, \eta)}{p(w_{1:N} | \alpha, \eta)} \quad (2)$$

However, the marginal probability $\mathrm{p}(w_N | \alpha, \eta)$ in the denominator, which is the probability of observing a particular set of documents given a particular $\alpha$ and $\eta$, is intractable to calculate exactly because it requires marginalizing over all possible instantiations of the latent distributions and there is a coupling between $\beta$ and $\theta$ [1,19]. In other words, it would require us to calculate:

$$p(w_N|\alpha,\eta) = \int\int p(\beta'|\eta)\, p(\theta'|\alpha) \prod_{n=1}^{N} p(z_n|\theta')\, p(w_n|z_n,\beta')\, d\beta' d\theta' \tag{3}$$

where the integration variables are all possible sets of topic-word distributions and document-topic distributions. Although this is intractable to calculate exactly, it can be approximated using sampling [1,8] or variation inference-based techniques [2]. Our chosen LDA library, Gensim, uses the latter approach.